\documentclass[12pt,preprint]{aastex}

\newcommand{\grs}      {\mbox{\rm\,GRS~1758--258}}
\newcommand{\onee}     {\mbox{\rm\,1E~1740.7--2942}}
\newcommand{\gx}       {\mbox{\rm\,GX~339--4}}
\newcommand{\cyg}      {\mbox{\rm\,Cyg~X--1}}

\begin{document}

\title{A Transition to the Soft State in \grs}

\author{D. M. Smith\altaffilmark{1},
W. A. Heindl\altaffilmark{2}, 
C. B. Markwardt\altaffilmark{3},
J. H. Swank\altaffilmark{3}}

\altaffiltext{1}{Space Sciences Laboratory, University of California Berkeley, 
Berkeley, CA 94720}
\altaffiltext{2}{Center for Astrophysics and Space Sciences, Code 0424, University
of California San Diego, La Jolla, CA 92093}
\altaffiltext{3}{NASA Goddard Space Flight Center, Code 666, Greenbelt, 
MD 20771}

\begin{abstract}

Near the end of 2001 February, the black-hole candidate (BHC) \grs\ made an
abrupt transition from a standard hard (low) state to a soft state.
Unlike \cyg\ and other BHCs, whose luminosity
increases during this transition, \grs\ was dimmer after the
transition.  We present observations with the Proportional Counter Array
on the \it Rossi X-ray Timing Explorer \rm and interpret the
phenomenon in the context of a ``dynamical'' soft state model.
Using this model we predicted that mass transfer from
the companion had ceased, and that the luminosity should decay
on a timescale of a few weeks.  The most recent data support this
prediction, being consistent with a decay time of 28 dy.
The current state is consistent with the ``off'' state of \grs\
reported by \it GRANAT/\rm Sigma in 1991-1992.

\end{abstract}


\section{Introduction}

\grs\ is one of only three black-hole candidates (BHCs) usually near its
maximum luminosity and usually in the hard state (the other two are
\cyg\ and \onee).  It displays shot-noise flickering and quasi-periodic
oscillations with characteristic timescales on the order of
1 s \citep{Sm97,Li00}.  It shows a core source and lobes in
the radio \citep{Ro92}, which caused it to be one of the first
sources referred to as a ``microquasar''.

The spectrum of \grs\ is generally a power law (with an exponential
cutoff beginning around 100 keV), occasionally with a weak thermal
component \citep{Me94,He98}.  Weekly observations with the
Proportional Counter Array (PCA) \citep{Ja96} on the \it Rossi X-ray
Timing Explorer (RXTE) \rm show that the power-law index can vary from
about 1.5 to 2.5, that the hard x-ray luminosity is always
dominated by the power law, and that the variability before 2001
(Figure 1) was not extreme \citep{Ma99,Sm01a}.  Earlier observations
by \it GRANAT/\rm Sigma at 40-150 keV showed the source
below detection level in Fall 1991 and Spring 1992 \citep{Gi93}, and
therefore more than an order of magnitude fainter than its typical \it
GRANAT \rm and \it RXTE \rm luminosity.  The Burst and Transient
Source Experiment (BATSE) on the \it Compton Gamma-Ray Observatory \rm
also saw the 20-100 keV flux from \grs\ vary dramatically \citep{Zh97c}.

\grs's spectral variations are not simultaneous with variations in its
luminosity, a newly-discovered characteristic which it shares with
\onee\ \citep{Ma99,Sm01a}.  In \cyg, the usual prototype of a BHC
system, there is a clearly defined progression from the hard to soft
state: the 2-10 keV flux increases dramatically as the spectrum
softens, the total x-ray luminosity increasing only slightly
\citep{Zh97a,Gi99}.  The softening takes two forms: the index of the
power law softens while at the same time the thermal component
brightens and goes to higher temperature.  In the transition back to
the hard state, the spectrum and luminosity also vary simultaneously.

\grs\ and \onee, in contrast, tend to be softest not when the 2-10 keV
flux is highest, but rather when the derivative of that count rate is
most negative \citep{Sm01a}.  We interpreted this behavior in terms of
a model by \citet{Ch95} in which BHC accretion operates in two
decoupled flows, a classical thin disk which produces thermal emission
and a nearly spherical, sub-Keplerian flow (``halo'' hereafter) which
scatters the thermal photons into the cut-off power law.  If mass
transfer from the companion suddenly dropped, altering both flows at
large radii, the change would propagate at the free-fall timescale in
the halo, but at the much slower viscous timescale in the thin disk.
Therefore, there would temporarily be an imbalance: the same number of
soft disk photons, but fewer hot halo electrons to scatter them.
Fewer scatters per photon softens the power law and, in addition, the
thinner halo would be more easily cooled by the disk photons.  In
\citet{Sm01a} we dubbed this a ``dynamical'' soft state and
hypothesized that it was responsible for almost all the softening
events observed in \grs\ and \onee, except for a brief period in Fall
of 2000 when \grs\ was at its brightest and seemed on the verge of
entering a true (``static'') soft state in which the halo self-cools
and collapses.

In this Letter we report on the recent transition of \grs\ to the
soft state, and interpret the data in terms of the dynamical soft
state model.

\section{Observations}

PCA snapshots of 1500 s were taken monthly of \grs\ in 1996, weekly
through 2000, and are being taken twice weekly from March 2001.  There
are no pointings for a period from November to January each year due
to a Sun-angle constraint on pointing.  Figure 1 shows the count rate
as a function of time in two energy bands: 2.5-4.0 keV and 10.0-25.0
keV.  Instrumental background and background due to Galactic Plane
diffuse emission \citep{Sm97,Ma99} have been subtracted.  The source
is kept off-axis to prevent contamination from GX 5-1 nearby.  All
results below use data from layer 1 of the PCA only.  The data were
analyzed with version 5.0.4 of HEAsoft.

\begin{figure}
\plotone{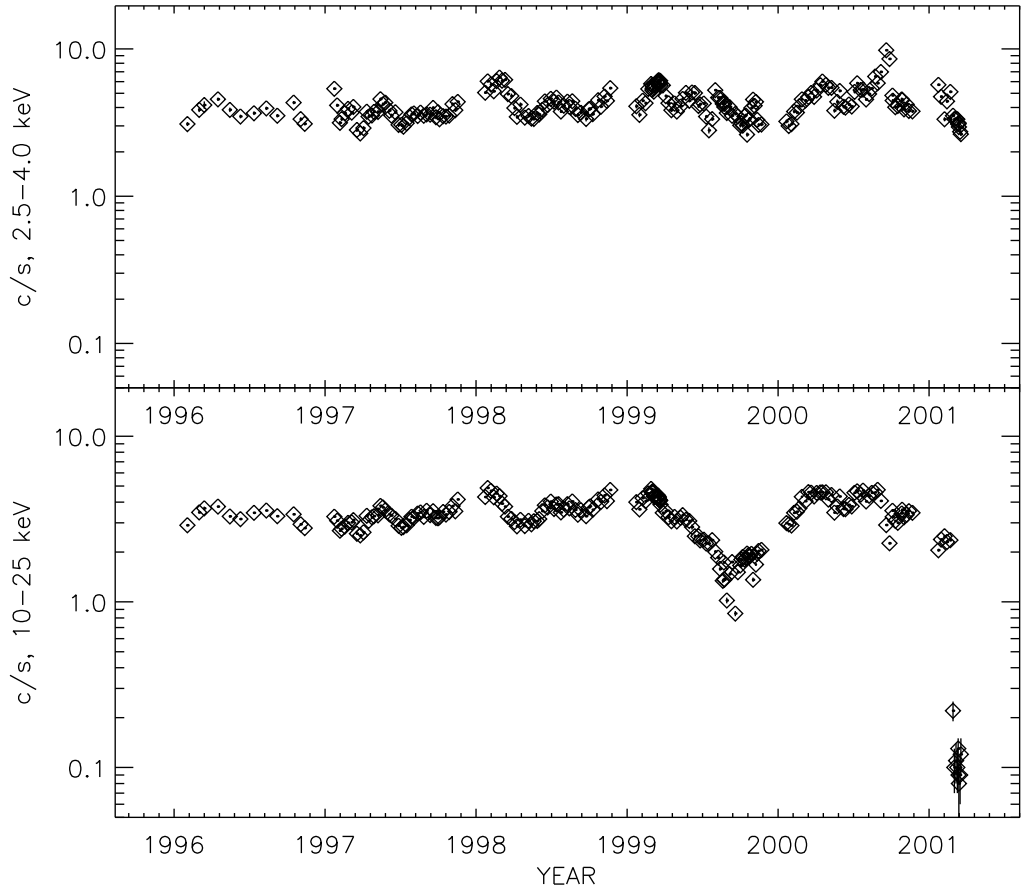}
\caption{PCA count rate from \grs\ vs. time in two energy bands.
These are raw count rates per PCU in layer 1.
A PCU (Proportional Counter Unit) is one of the PCA's five detectors.}
\end{figure}

There was a precipitous drop above 10 keV in late February of 2001
(Figure 1), but not in the soft band \citep{Sm01b,Sm01c}. On March
12-13, 31 ksec of public observations were made with \it RXTE \rm in
three separate intervals.  Since the new soft state would have been
undetectable to Sigma, it is quite possible that this was the state
that instrument observed in 1991-1992.

\section{Results}

\begin{figure}
\plotone{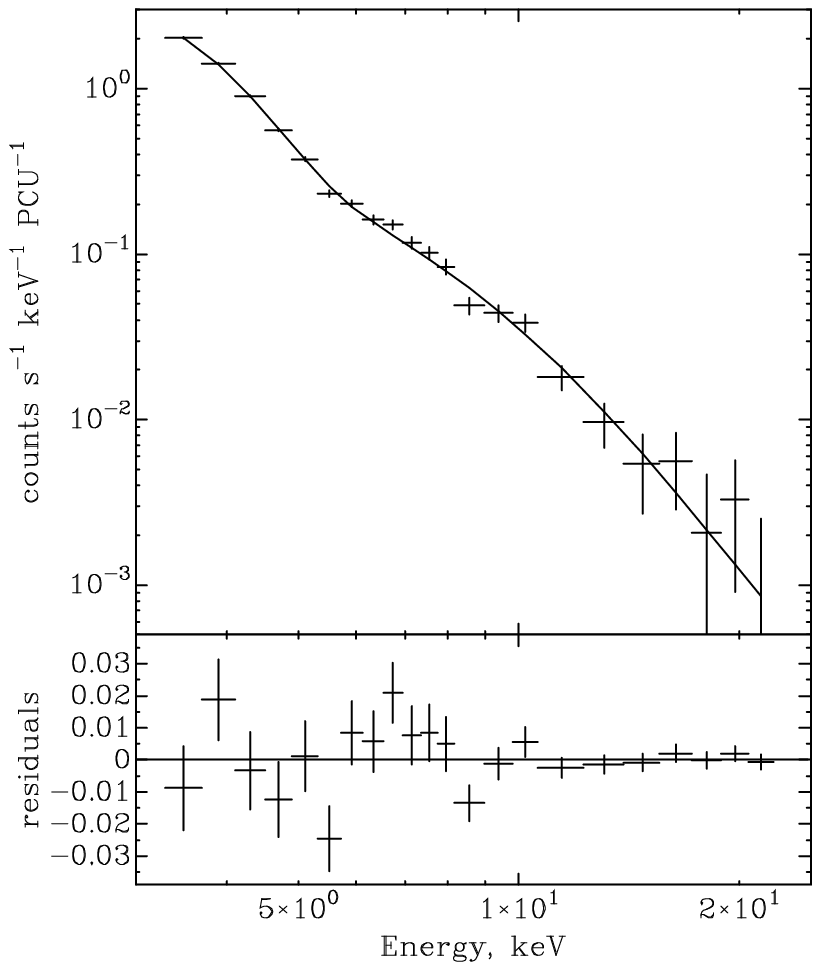}
\caption{Upper panel: count spectrum and power-law-plus-blackbody 
fit for the deep PCA pointing in the soft state of \grs.  For display,
the data from all four PCUs and all three parts of the deep pointing
are summed.  The fits used to derive parameters mentioned in the text
were simultaneous fits to independent data sets representing each PCU
in each part of the deep pointing.  Lower panel: residuals between the
data and the fit.}
\end{figure}

\begin{figure}
\plotone{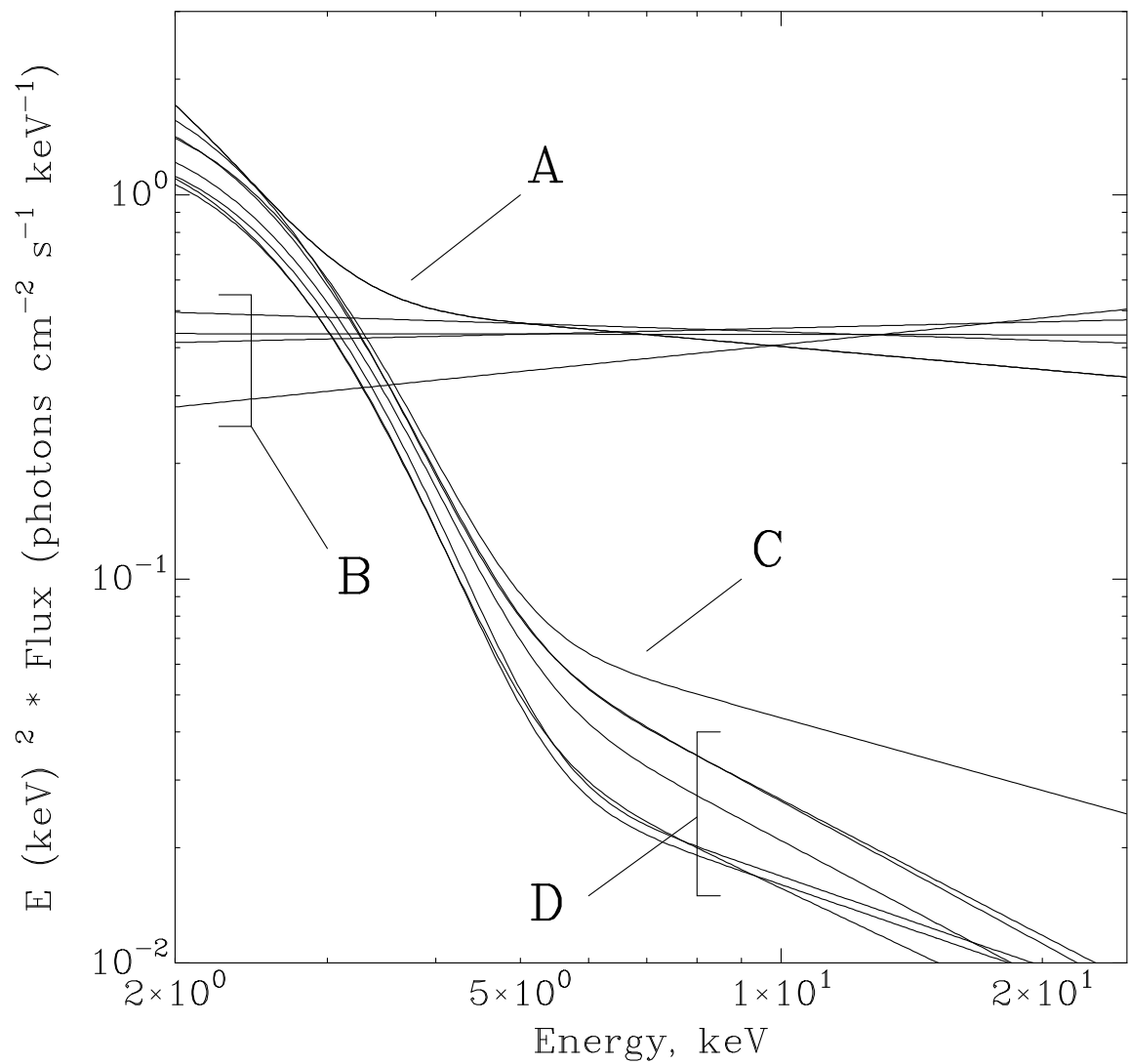}
\caption{ The fitted spectral models from January 23 to March
18. ``A'': January 23.  ``B'': January 29 to February 21. ``C'':
February 27.  ``D'': March 2 to March 18.  The effect of interstellar
absorption has been removed. }
\end{figure}

\begin{figure}
\plotone{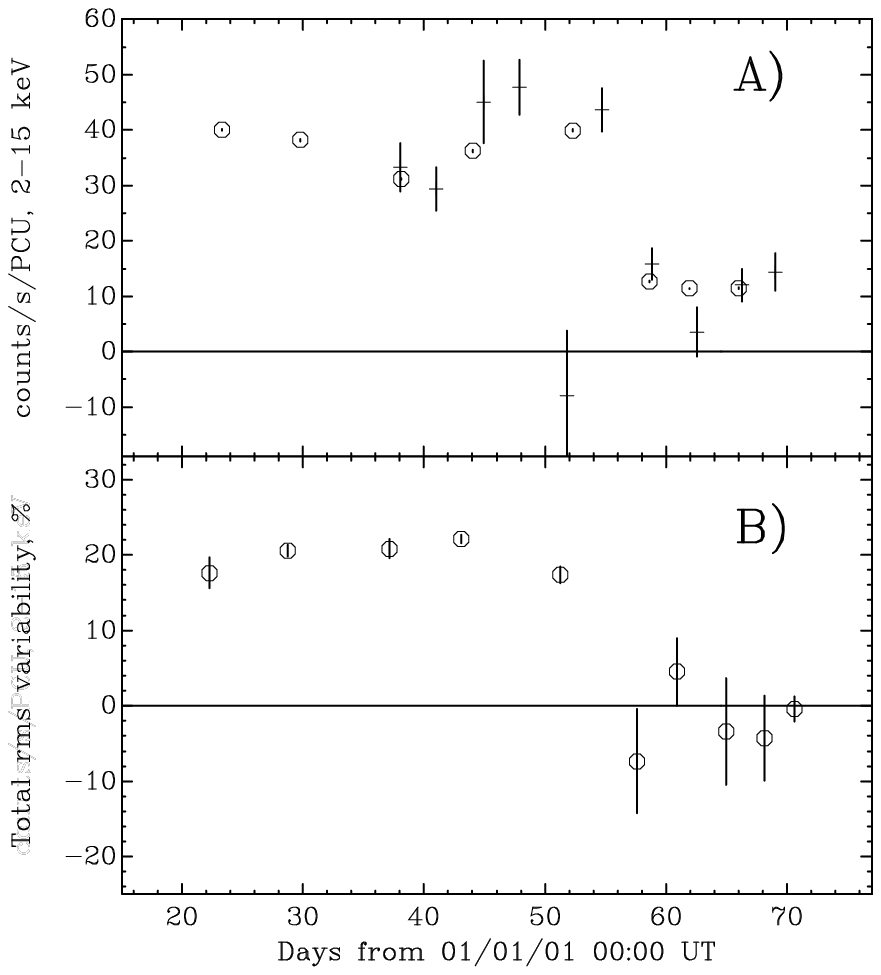}
\caption{
Evolution of count rate and rms variability in \grs\ around the state
transition.  A) PCA count rate from 2-15 keV, showing both the
monitoring data of Figure 1 (circles -- the error bars are almost too
small to see) and data from the Galactic-bulge scans (crosses -- see
text).  Note the brief drop to low flux around day 51 (February
20). B) Total rms variability from 0.004-16 Hz and 2-10 keV.  Poisson
noise has been subtracted.}
\end{figure}

Figure 2 shows the PCA spectrum of the combined March 12-13 data,
including the residuals from a fit to a power law plus a blackbody.
Because the PCA fits begin at 3 keV, the fitted hydrogen absorption column and
blackbody intensity are strongly correlated and can't both be
determined well.  We therefore fixed the column at
the ASCA value \citep{Me97}: 1.5 $\times 10^{22}$ atoms
cm$^{-2}$.  The fitted blackbody temperature was (0.395 $\pm$ 0.006)
keV and the power-law photon index was (2.89 $\pm$ 0.12).  Using a
multicolor disk blackbody, the maximum temperature was (0.464 $\pm$
0.007) keV and the power law index (2.75 $\pm$ 0.12).
The blackbody luminosity ($3.0\times 10^{-9}$ ergs
cm$^{-2}$s$^{-1}$, or $2.6\times 10^{37}$ ergs s$^{-1}$ at 8.5 kpc) is
the highest seen with \it RXTE \rm, although only about 35\% higher
than in some spectra from near the peak in the top panel of Figure 1
in late 2000. 

As is apparent from the residuals in Figure 2, the addition of an iron
K-shell line improves the fit.  Fixing the energy at 6.7 keV and
the width at 0.1 keV (narrower than the PCA resolution), the derived
equivalent width is 300 eV for both the blackbody and disk blackbody
models.  The interpretation of a weak, instrumentally broad line near
the crossover point of two continuum components is problematic, and we
do not claim this as a certain detection.  No significant improvement
in $\chi^{2}$ resulted from letting the energy or width of the line
vary.

Figure 3 shows the fits to the monitoring observations in early 2001,
before and after the state transition.  The first observation of 2001
(January 23, labeled ``A'' in Figure 3) displays both a bright power law and
a significant blackbody peak (this is also occasionally seen at other
periods from 1996-2000).  The soft excess from 1.0-2.4 keV
measured by \citet{Me94} with ROSAT in 1994 was
2.8$\times 10^{-10}$ ergs cm$^{-2}$s$^{-1}$, about half the value for
this spectrum (5.7$\times 10^{-10}$ ergs cm$^{-2}$s$^{-1}$ from
1.0-2.4 keV, or 7.0$\times 10^{-10}$ ergs cm$^{-2}$s$^{-1}$
bolometric).  The remaining spectra before the transition have no
detectable blackbody and a harder power-law index.  The first spectrum
after the transition (labeled ``C'') has somewhat more of a residual
power-law tail than the spectra after it.

Figure 4a shows a combined lightcurve around the time of the
transition, using data from two sources: the monitoring campaign
discussed above, and a second \it RXTE \rm campaign of twice-weekly
scans of the whole Galactic-bulge region \citep{Mar99,Mar01}.  For
each scan, the data are fitted to a model of point sources and diffuse
emission.  One scan point on February 20 implies that the source may
have dropped to a low level in less than 3.9 dy and then rebounded
completely less than 11.3 hr later.  Generally speaking the pointed
observations and bulge-scan intensities agree to within $\sim2\sigma$.
However, the proximity of \grs\ to the bright source GX~5--1 may
potentially introduce unmodelled systematic errors into the scan data.

Figure 4b shows the evolution of the root mean square (rms) variability
beyond Poisson noise.  As is typical for BHC soft states
\citep[e.g.][]{Ma84}, the variability drops dramatically.
The last data point represents all the deep-pointing data
from March 12-13.  At this time the rms variability (0.004-16 Hz, 2-10
keV) was consistent with zero, with a 3$\sigma$ upper limit of 2.5\%.

\section{Discussion}

There are similarities and differences between this hard-to-soft
transition and transitions in other BHCs.  One of the best-monitored
transitions was in \cyg\ in 1996.  The 1.3-200 keV luminosity as
measured by the \it RXTE \rm All-Sky Monitor (ASM) and BATSE,
corrected for interstellar absorption, jumped
upwards by about 35\% \citep{Zh97a}.  To compare, we extrapolated our
spectral fits to this energy range and removed the interstellar
absorption using the fixed column density discussed above.  \grs\ and
other BHCs have an exponential cutoff in the hard state beyond the PCA
range, and we used cutoff parameters from a deep \it RXTE \rm pointing
to \grs\ in 1996 \citep{He98}.  We find that the extrapolated 1.3-200
keV luminosity of \grs\ dropped in the hard-to-soft transition by
about the same amount that it rose in \cyg\ in 1996.

That soft state in \cyg, however, was not typical of BHC
soft states in general.  The power-law flux at 25 keV
dropped only by a factor of 3 from the hard state \citep{Zh97b}
compared to the factor of 50 apparent in Figure 3, and
there was still 20\% rms of fast variability \citep{Cu97}.
This led \citet{Be96} to conclude that this ``soft'' state 
should have been called an ``intermediate'' state.
If the hard/intermediate/soft progression is 
a monotonic increase in accretion rate as generally thought,
one would expect that the more ``complete'' transition
in \grs\ would result in an even larger increase in luminosity,
not a decrease.

\gx\ entered the soft state in 1990 and 1998 with similar spectral and
timing characteristics to the new soft state in \grs.  Between April
and August of 1990, the power-law tail in \gx\ dropped by at least an
order of magnitude at 10 keV and the rms variability in the soft state
was only (6.1 $\pm$ 2.7)\% \citep{Ge91}.  For the 1998 transition,
\citet{Be99} compared the 2.5-20 keV unabsorbed flux in the soft state
to data taken in the hard state several months previously.  They found
that the total flux in this band increased by about a factor of two.
We can evaluate our PCA data in this band without extrapolation.  We
find again that, contrary to the behavior of the other BHC, \grs\ saw
a decrease in absorption-corrected flux in the energy band chosen for
comparison, in this case a drop of a factor of 3.3 from 2.5-20 keV
between February 21 and March 2.

BATSE was the first to observe a soft state in \onee, with a power law
index of -2.6 from 20-100 keV \citep{Zh97c}.  The soft spectrum
crossed the usual harder spectrum at 20 keV, so again there would have
been an increase rather than a drop in the power-law component over
the 3-25 keV band.

\begin{figure}
\plotone{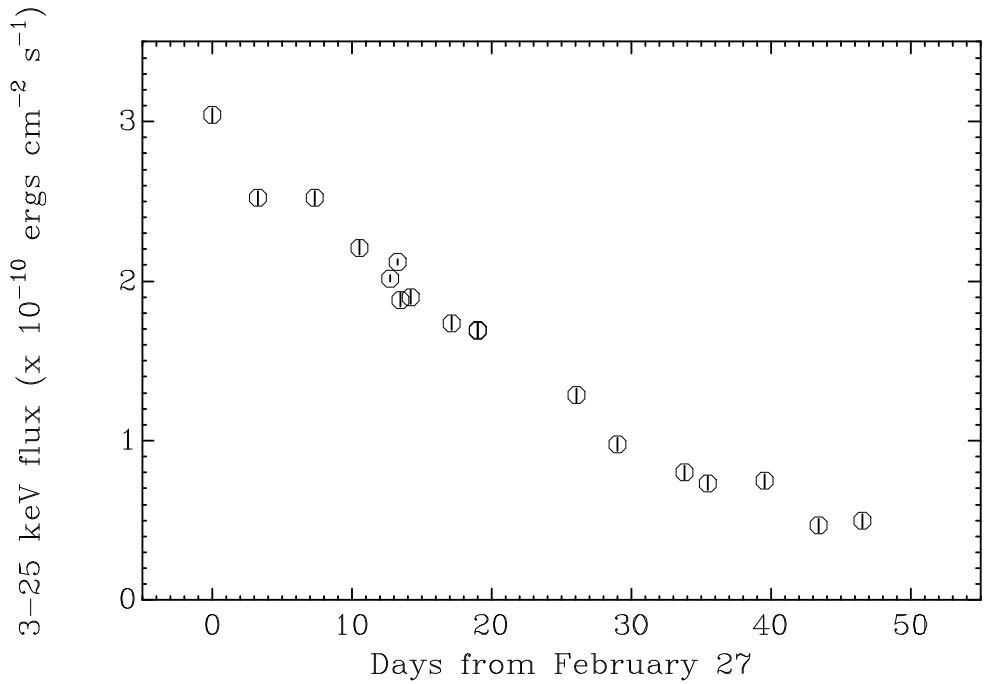}
\caption{Energy flux in \grs, 3-25 keV, as a function of time since
the state transition on February 27, uncorrected for interstellar
absorption.  Note that the scatter of the data is much greater than
the error bars, which are shown, so that the quality of a fit to any
smooth function is poor (reduced $\chi^{2} \sim 10$).  }
\end{figure}

While the transitions in \cyg, \gx, and other BHCs are probably indeed
caused by an increase in accretion causing self-cooling and partial
collapse of the halo (static soft state), we suggest that this
transition in \grs\ is instead caused by the sudden shutoff of all, or
nearly all, of the mass transfer from the compact object's companion.
This is the extreme instance of the dynamical soft state described in
\citet{Sm01a}.  Given the hypothesis \citep{Ch95} that the disk and
halo flows are independently fed directly from the companion, the halo
would then vanish immediately leaving the disk to decay away on a
viscous timescale on the order of a month or more \citep{Ma99,Sm01a}.
We wouldn't expect \cyg\ to show a dynamical soft state, because
as a wind accretor its disk is expected to be small, and therefore
have a short viscous time \citep{Sm01a}.

It is possible that, beneath the halo, the disk emission was more or
less constant throughout early 2001, and when it wasn't seen, the halo
was optically thick.  In the first spectrum of the year (``A'' in
Figure 3), with a blackbody luminosity about a factor of 4 below the
naked disk, the optical depth of the halo would then have been near
unity.  The changes in the slope of the power law support this
interpretation: the slope is hardest when the halo is assumed to be
thickest, and repeated scatterings should produce a harder power law
\citep{Sh76}.  If the apparent drop on February 20 (Figure 4a) is
genuine, then the time for the halo to re-establish itself would be
less than 11.3 hr, which is still longer than the free-fall time from
the outer disk.

Viewing the disk and halo flows as independent, and assuming that
the disk flow did not change as the halo disappeared, the loss of fast
variability in the transition implies that the flickering commonly
observed in the hard state of BHCs is either intrinsic to the halo
flow alone, or else due to an interaction at the boundary between the
dense, radially slow-moving disk and the thin, radially fast-moving
halo.

When the transition to the soft state occurred, we predicted
\citep{Sm01c}, based on the dynamical picture and the disk timescales
we previously derived \citep{Ma99,Sm01a}, that mass transfer from the
companion had stopped and the soft-state emission would decay away
with a timescale of a few weeks.  Figure 5 shows the flux (uncorrected
for absorption) from 3-25 keV as a function of time starting with the
first soft-state spectrum (February 27) to the most recent data
available (April 15).  The shape is neither exponential or linear, but
somewhere in between.  The best fit exponential gives a (27.9 $\pm$
1.7) dy time constant.  The latest fluxes approach the limit of the
systematic uncertainty in the Galactic diffuse emission.

If there is still a small mass transfer from the companion, the
spectrum may eventually make a transition to the pure, very faint hard
power law seen in \gx\ in its lowest luminosity (``off'') state
\citep{Ma84,Il86,Men97,As98,Ko00}.  Observations of \gx\ were
never taken immediately following the drop from higher fluxes, and
it may show a dynamical soft state at those times.  Eventually, our
understanding of the state change in \grs\ may apply to the state
changes in the soft x-ray transients (the outbursts of the more
numerous BHCs without persistent emission), which can show hysteresis
with a hard-to-soft transition near peak flux followed by a much later
soft-to-hard transition at much lower flux \citep{Mi95}.


\begin{thebibliography}{}

\bibitem[Asai et al.(1998)]{As98}Asai, K., Dotani, T., Hoshi, R., Tanaka, Y., Robinson, C. R., 
\& Terada, K., 1998, PASJ, 50, 611

\bibitem[Belloni et al.(1999)]{Be99}Belloni, T., Mendez, M.,
van der Klis, M., Lewin, W. H. G., \& Dieters, S.
1999, \apjl, 519, L159

\bibitem[Belloni et al.(1996)]{Be96}Belloni, T., Mendez, M.,
van der Klis, M., Hasinger, G., Lewin, W. H. G., \& van Paradijs, J.
1996, \apjl, 472, L107

\bibitem[Chakrabarti \& Titarchuk(1995)]{Ch95} Chakrabarti, S. K. \& Titarchuk, L. G. 1995, \apj, 455, 623

\bibitem[Cui et al.(1997)]{Cu97}Cui, W., Zhang, S. N., Focke, W., \& Swank, J. H. 1997, \apj, 484, 383

\bibitem[Gierli\'{n}ski et al.(1999)]{Gi99}Gierli\'{n}ski, M., Zdziarski, A. A., Poutanen, J., Coppi, P. S., Ebisawa, K., \& Johnson, W. N. 1999,
MNRAS, 309, 496

\bibitem[Gilfanov et al.(1993)]{Gi93}Gilfanov, M. et al. 1993, \apj, 418, 844

\bibitem[Grebenev et al.(1991)]{Ge91}Grebenev, S. A.,
Syunyaev, R. A., Pavlinskii, M. N., \& Dekhanov, I. A. 1991,
Soviet Astr. Lett. 17, 413

\bibitem[Heindl \& Smith(1998)]{He98}Heindl, W. A., \& Smith, D. M. 1998, \apjl
506, L35

\bibitem[Ilovaisky et al.(1986)]{Il86}Ilovaisky, S. A., Chevalier, C., Motch, C., 
\& Chiappetti, L., 1986, \aap, 164, 67

\bibitem[Jahoda et al.(1996)]{Ja96}Jahoda, K. et al. 1996, SPIE, 2808, 59

\bibitem[Kong et al.(2000)]{Ko00}Kong, A. K. H., Kuulkers, E., Charles, P. A., \& Homer, L. 2000,
MNRAS, 312, L49

\bibitem[Lin et al.(2000)]{Li00}Lin, D., et al. 2000, \apj, 532, 548

\bibitem[Maejima et al.(1984)]{Ma84}Maejima, Y., Makishima, K., Matsuoka, M.,
Ogawara, Y., Oda, M., Tawara, Y., \& Doi, K. 1984, \apj, 285, 712

\bibitem[Main et al.(1999)]{Ma99}Main, D. S., Smith, D. M., Heindl, W. A., Swank, J. H., Leventhal, M., Mirabel, I. F., \&
Rodr\'{i}guez, L. F. 1999, \apj, 525, 901

\bibitem[Markwardt et al.(1999)]{Mar99}Markwardt, C. B., Marshall, F. E., Swank, J. H. \& in't Zand, J. J. M. 1999, BAAS, 31, 970

\bibitem[Markwardt et al.(2001)]{Mar01}Markwardt, C. B. et al. 2001,
in preparation

\bibitem[Mendez \& van der Klis(1997)]{Men97}Mendez, M., \& van der Klis, M., 1997, \apj, 479, 926

\bibitem[Mereghetti et al.(1994)]{Me94}Mereghetti, S., Beloni, T., \& Goldwurm, A. 1994, \apjl, 433, L21

\bibitem[Mereghetti et al.(1997)]{Me97}Mereghetti, S., Cremonesi, D. I., Haardt, F., Murakami, T.,
Beloni, T., \& Goldwurm, A. 1997, \apj, 476, 829

\bibitem[Miyamoto et al.(1995)]{Mi95}Miyamoto, S., Kitamoto, S., Hayashida, K.,
\& Egoshi, W. 1995, \apjl 442, L13

\bibitem[Rodriguez et al.(1992)]{Ro92}Rodriguez, L. F., Mirabel, I. F., \& Marti, J. 1992, \apjl, 401, L15

\bibitem[Shapiro et al.(1976)]{Sh76}Shapiro, S., Lightman, A., \& Eardley, D. 1976, \apj, 204, 187

\bibitem[Smith et al.(1997)]{Sm97}Smith, D. M., Heindl, W. A., Swank, J. H., Leventhal, M.,
Mirabel, I. F., \& Rodriguez, L. F. 1997, \apj, 489, L51

\bibitem[Smith et al.(2001a)]{Sm01a}Smith, D. M., Heindl, W. A., \& Swank, J. H.,
2001, submitted to \apj, astro-ph/0103304

\bibitem[Smith et al.(2001b)]{Sm01b}Smith, D. M., Markwardt, C. B., Heindl, W. A., \& Swank, J. H.,
2001, IAUC \#7595

\bibitem[Smith et al.(2001c)]{Sm01c}Smith, D. M., Heindl, W. A., Markwardt, C. B., \& Swank, J. H.,
2001, ATEL \#66

\bibitem[Zhang et al.(1997a)]{Zh97a}Zhang, S. N., Cui, W., Harmon, B. A., Paciesas, W. S., Remillard, R. E., \& van Paradijs, J.
1997, \apj, 477, 95

\bibitem[Zhang et al.(1997b)]{Zh97b}Zhang, S. N., Cui, W., Harmon, B. A., \& Paciesas, W. S. 1997, AIP Conf. Proc., 410, 839

\bibitem[Zhang et al.(1997c)]{Zh97c}Zhang, S. N., Harmon, B. A., \& Liang, E. P. 1997, AIP Conf. Proc., 410, 873

\end{thebibliography}
\end{document}